\documentstyle[sprocl]{article}

\def\be{\begin{equation}}
\def\ee{\end{equation}}
\def\bea{\begin{eqnarray}}
\def\eea{\end{eqnarray}}

\begin{document}

\title{HIGHER--TWIST CONTRIBUTIONS TO $F_L$ AND $F_2$ FROM 
INSTANTONS${}^\ast$}
\footnotetext{${}^\ast$ Presented at DIS 2000, Liverpool, 
April 25 -- 30, 2000.}

\author{B. DRESSLER\footnote{E-mail: birgitd@tp2.ruhr-uni-bochum.de}, 
C. WEISS\footnote{E-mail: weiss@tp2.ruhr-uni-bochum.de}}

\address{Institut f\"ur Theoretische Physik II,
Ruhr--Universit\"at Bochum, \\ D--44780 Bochum, Germany} 

\author{M. MAUL\footnote{E-mail: maul@thep.lu.se}}

\address{Department of Theoretical Physics, Lund University,\\
S\"olvegatan 14A, S - 223 62 Lund, Sweden}


\maketitle\abstracts{We study the twist--4 contribution to
the unpolarized nucleon structure functions, $F_L$ and $F_2$, 
in an approach where the vacuum structure of QCD is described 
by a dilute medium of instantons ($\bar\rho / \bar R \ll 1$). 
The dominant power corrections come from the twist--4 quark--gluon 
operators and are of the order $1/ (\bar\rho^2 Q^2)$.
The contributions from four--quark operators are suppressed.
The approach predicts large $1/Q^2$--corrections to $F_L$, in 
agreement with the results of QCD fits to 
the data.}

The operator product expansion of QCD explains not only the scaling
behavior of structure functions at asymptotically large $Q^2$ ---
including the logarithmic corrections, which can be computed using
perturbation theory --- but allows to classify also the power ($1/Q^2$--) 
corrections to scaling. The latter are determined by the
expectation values of certain operators of non-leading twist ($>
2$) in the target hadron state \cite{Shuryak:1982kj}. While a fully 
quantitative theory combining logarithmic and power corrections is 
presently out of reach, it is interesting to study these higher--twist matrix 
elements in their own right. These are ``universal'' measures of the
non-perturbative correlations of the quark and gluon fields inside
hadrons, which appear also in other contexts, {\it e.g.} in the description 
of ``intrinsic'' heavy quark contributions to light hadron 
properties~\cite{Polyakov:1999rb}. Knowledge of these matrix elements 
would be useful 
for a more precise extraction of the leading--twist parton distributions 
from the DIS data, and for a better understanding of 
parton--hadron duality \cite{Ji:1995br}.
\par
Two basic types of twist--4 operators appear in the
$1/Q^2$--power corrections to the second moments of the unpolarized 
structure functions (for details, see 
Refs.\cite{Shuryak:1982kj,Dressler:1999zi}). One are quark--gluon operators, 
which describe the influence in DIS of the non-perturbative gluon field in 
the target on the ``struck'' quark:
\begin{equation}
\left.
\begin{array}{rcr}
\langle p| \, \bar\psi_f \gamma_{\alpha} (D^\gamma 
F_{\gamma\beta}) \psi_f \, |p \rangle \; - \; \mbox{trace} 
& \, = \, & 2 \, A_f 
\\[1.5ex]
\langle p| \, \bar\psi_f \gamma^\gamma \gamma_5 i \left(
\nabla_{\alpha} \widetilde{F}_{\beta\gamma } 
+ \widetilde{F}_{\alpha\gamma} \nabla_{\beta} 
\right) \psi_f \,
|p \rangle  \; - \; \mbox{trace}
& \, = \, & 2 \, B_f 
\end{array}
\right\}
\times \left( p_\alpha p_\beta - \frac{p^2}{4} g_{\alpha\beta} \right) .
\label{A_B_def} 
\end{equation}
Here, $F_{\alpha\beta} \, (\widetilde F_{\alpha\beta})$ denote the (dual)
gluon field.
The other are four--quark operators, describing the interference contribution
to DIS from scattering off two different quarks in the target:
\be
g^2 \, \langle p | \, 
\bar\psi_f \frac{\lambda^a}{2} \gamma_\alpha\gamma_5 \psi_f \;
\bar\psi_{f'} 
\frac{\lambda^a}{2} \gamma_\beta\gamma_5 \psi_{f'} \,
| p \rangle \;\; = \;\; 2 \, C_{ff'} 
\left( p_\alpha p_\beta - \frac{p^2}{4} g_{\alpha\beta} \right) .
\label{C_def}
\ee
In this context two questions arise:
\begin{itemize}
\item 
What are the relevant non-perturbative scales determining the values
of the twist--4 matrix elements, $A, B$ and $C$?
\item
Are the matrix elements of both types of operators of equal magnitude
or is one of them dominant?
\end{itemize}
\par
It is interesting to address these questions from the point of view of the 
instanton vacuum, where one studies
non-perturbative effects generated by a dilute ``medium'' of instantons
and antiinstantons. These vacuum fluctuations are known to be 
responsible for the spontaneous breaking of the chiral symmetry of QCD and a 
variety of other effects; see \cite{Schafer:1998wv} for a review. The 
properties of the
instanton ``medium'' have been derived within a variational approach
to the QCD ground state \cite{Diakonov:1984hh}, and have also been 
extensively studied in lattice simulations \cite{Schafer:1998wv}.
A crucial property is that the average size of the instantons
in the medium, $\bar\rho = (600 \, {\rm MeV})^{-1}$, is small compared
to their average distance, $\bar R$, with $\bar\rho / \bar R \approx 1/3$
(``diluteness''). This ratio provides a small parameter which allows to 
classify non-perturbative effects generated by the instanton medium.
For instance, the dynamical quark mass, $M$, which appears in the 
spontaneous breaking of chiral symmetry, is parametrically of the order
$M \sim (\bar\rho / \bar R )^2 \, \bar\rho^{-1}$. As a result, {\it e.g.}\ 
the pion weak decay constant is parametrically suppressed:
\begin{equation}
F_\pi^2 \;\; \sim \;\; \int\limits_{k^2 < \bar\rho^{-2}} \!
\frac{d^4 k}{(2 \pi )^4} \frac{M^2}{(k^2 + M^2)^2}
\;\; \sim \;\; M^2 \log M\bar\rho
\;\; \sim \;\; \left(\frac{\bar\rho}{\bar R}\right)^4 
\log \left(\frac{\bar\rho}{\bar R}\right) \bar\rho^{-2}.
\label{fpi}
\end{equation}
Similar parametric estimates can be performed for other observables.
\par
Following the same philosophy we can estimate also the magnitude of the 
twist--4 nucleon matrix elements, (\ref{A_B_def}) and 
(\ref{C_def}).\cite{Dressler:1999zi,Balla:1998hf}
When computing the matrix element of the twist--4 quark--gluon operators,
(\ref{A_B_def}), the gauge 
field in the operator is replaced by the field of a single (anti--) 
instanton, which then interacts 
with the fermion fields in the operator through its fermionic zero modes.
(The same zero--mode induced interaction is instrumental in the dynamical 
breaking of chiral symmetry.) One finds that the isoscalar nucleon
matrix element is determined by an integral of the form
\begin{equation}
B_u + B_d \;\; \sim \;\; \int\limits_{k^2 < \bar\rho^{-2}}
\frac{d^4 k}{(2 \pi )^4} \frac{k^2}{(k^2 + M^2)^2}
\;\; \sim \;\; \bar\rho^{-2} .
\label{B_res}
\end{equation}
Contrary to the pion decay constant, (\ref{fpi}), this integral is 
quadratically divergent, so that its value is determined by the square of 
the ultraviolet cutoff --- the inverse instanton size, $\bar\rho^{-1}$.
A more quantitative estimate gives $B_u + B_d \, \approx \,  
0.9 \, \bar\rho^{-2} \, = \, (570 \, {\rm MeV})^2$.
The matrix element has a large numerical value, because of the smallness of
the average instanton size. The corresponding matrix element of the first
operator, (\ref{A_B_def}), is accidentally suppressed, since 
$D^\gamma F_{\gamma\beta} \equiv 0$ in the 
field of one instanton. For the matrix elements of the four--fermionic
operators, on the other hand, we find
\begin{equation}
C_{uu}, C_{dd}, C_{ud} \;\; \sim \;\; F_\pi^2 \;\;
\;\; \sim \;\; \left(\frac{\bar\rho}{\bar R}\right)^4 
\log \left(\frac{\bar\rho}{\bar R}\right) \bar\rho^{-2}.
\end{equation}
These matrix elements are parametrically suppressed relative to that of the 
operator $B$, (\ref{B_res}). In addition, they are suppressed in
the coupling constant, $g^2$. Numerically, a value of 
$(50\, {\rm MeV})^2$ for the isoscalar matrix element was found in
Ref.\cite{Dressler:1999zi} --- two orders
of magnitude smaller than for the quark--gluon operator!
\par 
To summarize, the parametric order of the higher--twist matrix elements 
obtained from the instanton vacuum implies that:
\begin{itemize}
\item 
Matrix elements of twist--4 four--quark operators, (\ref{C_def}), are 
parametrically (and numerically) small compared to those of quark--gluon 
operators, (\ref{A_B_def}).
\item
The relevant scale for twist--4 quark--gluon operators is the square of the
average instanton size, $\bar\rho^{-2} = (600 \, {\rm MeV})^2$. The dominant
power corrections are thus expected to be of the order $1/(\bar\rho^2 Q^2)$.
\end{itemize}
These findings are in agreement with the results of QCD fits to the data 
for $F_L$ and $F_2$, which show a sizable $1/Q^2$ corrections to $F_L$ 
consistent with (\ref{B_res}). \cite{Dressler:1999zi}
\end{document}